\documentclass{article}
\usepackage{amsmath}
\usepackage{amsfonts}
\usepackage{amssymb}

\begin{document}

\title{\textbf{Higher regularity properties of mappings and morphisms}}
\author{\textbf{Steven Duplij}\thanks{ On leave of absence from Kharkov National University,
Kharkov 61001, Ukraine} \thanks{ E-mail: Steven.A.Duplij@univer.kharkov.ua}
\thanks{ Internet: http://gluon.physik.uni-kl.de/\~{}duplij} \hskip0.7em and
\textbf{W\l adys\l aw Marcinek}\thanks{ E-mail: wmar@ift.univ.wroc.pl}\\
Institute of Theoretical Physics, University of Wroc\l aw,\\Pl. Maxa Borna
9, 50-204 Wroc{\l}aw,\\Poland}
\date{18 March 2000}
\maketitle
\begin{abstract}
We propose to extend ``invertibility'' to ``regularity'' for categories in
general abstract algebraic manner. Higher regularity conditions and
``semicommutative'' diagrams are introduced. Distinction between commutative
and ``semicommutative'' cases is measured by non-zero obstruction proportional
to the difference of some self-mappings (obstructors) $e^{\left(  n\right)  }$
from the identity. This allows us to generalize the notion of functor and to
``regularize'' braidings and related structures in monoidal categories. A
``noninvertible'' analog of the Yang-Baxter equation is proposed.
\end{abstract}

\newpage

Generalizing transition from invertibility to regularity is a widely used
method of abstract extension of various algebraic structures. The idea of
regularity was firstly introduced by von Neumann \cite{neu} and applied by
Penrose to matrices \cite{pen1}. Then intensive study of regularity was
developed in many different fields, e.g. generalized inverses theory
\cite{rabson,rao/mit,nashed} and semigroup theory
\cite{mun/pen,cli5,lal1,howie,lawson}. In this paper we consider this concept
in categorical language \cite{mitchell,maclane1} and consider higher
regularity conditions (firstly introduced in the noninvertible extension of
supermanifold theory \cite{dup14,dup18}) in the most abstract form.

\section{Noninvertible mappings and morphisms}

Here we introduce general mappings (and later morphisms) without usual
requirement of ``invertibility''. We start from some standard well-known facts
(see e.g. \cite{cohn}). Let $X$ and $Y$ be two arbitrary sets. A mapping $f$
from $X$ to $Y$ is defined by a prescription which assigns an element of $Y$
to each element of $X$, i.e. $f:X\rightarrow Y$. Injective mapping (injection)
assigns different images to different elements, and in surjective mapping
(surjection) every image has at least one pre-image. Bijection has both
properties. Usually inverse mapping $f^{-1}$ is defined as a new mapping
$g:Y\rightarrow X$ which assigns to each $y\in Y$ such $x\in X$ that $f\left(
x\right)  =y$ and so $f^{-1}=g$. For injective $f$ and any $A\subset X$ it is
imposed the following ``invertibility'' condition

\begin{equation}
f^{-1}\left(  f\left(  A\right)  \right)  =A.\label{1}
\end{equation}
For surjective $f$ and $B\subset Y$ the standard ``invertibility'' condition is

\begin{equation}
f\left(  f^{-1}\left(  B\right)  \right)  =B.\label{2}
\end{equation}

These conditions are very strong, because they imply possibility to solve the
equation $f\left(  x\right)  =y$ for all elements. This situation is very much
artificial and takes place in only obvious special cases, when functions are
defined by invertible operations. Therefore abstract consideration of quantum
field theory constructions is usually restricted by group theory methods. But
in many cases, especially while considering supersymmetric theories, there
naturally appear noninvertible morphisms \cite{dup14,dup18} and semigroups
\cite{dup6,dup11,dup-hab}. That obviously needs extending some first
principles and assumptions.

We propose to extend the ``invertibility'' conditions (\ref{1})--(\ref{2}) in
the following way (which comes from analogy of regularity in semigroup theory
\cite{howie,petrich3}). Instead of $f^{-1}$ mapping we introduce less
restricted ``regular'' $f^{\ast}$ mapping by extending ``invertibility'' to
``regularity'' in following way

\begin{equation}
f\left(  f^{\ast}\left(  f\left(  A\right)  \right)  \right)  =f\left(
A\right)  .\label{1r}
\end{equation}

For the second equation (\ref{2}) we have the ``reflexive regularity'' condition

\begin{equation}
f^{\ast}\left(  f\left(  f^{\ast}\left(  B\right)  \right)  \right)  =f^{\ast
}\left(  B\right)  .\label{2r}
\end{equation}

\section{Invertibility and regularity of morphisms}

Among all mappings we distinguish morphisms satisfying closure and
associativity, which defines a category $\frak{C}$ with objects
$\operatorname*{Ob} \frak{C}$ as sets $X,Y,Z$ and morphisms
$\operatorname*{Mor}\frak{C}$ as mappings $f:X\rightarrow Y$ between them (or
$f=\operatorname*{Mor}\left(  X,Y\right)  $). For composition $h:X\rightarrow
Z$ of morphisms $f:X\rightarrow Y$ and $g:Y\rightarrow Z$ instead of $h\left(
x\right)  =g\left(  f\left(  x\right)  \right)  $ for mappings we use the
notation $h=g\circ f$. Associativity implies that $h\circ\left(  g\circ
f\right)  =\left(  h\circ g\right)  \circ f=h\circ g\circ f$. Right
cancellative morphisms are \textit{epimorphisms} which satisfy $g_{1}\circ
f=g_{2}\circ f\Longrightarrow$ $g_{1}=g_{2}$, where $g_{1,.2}:Y\rightarrow Z$
and left cancellative morphisms are \textit{ monomorphisms} which satisfy
$f\circ h_{1}=f\circ h_{2}\Longrightarrow$ $h_{1}=h_{2}$, where $h_{1,.2}
:Z\rightarrow X$.

Let us consider ``invertibility'' properties of morphisms in general. If $f$
satisfies the ``right invertibility'' condition $f\circ f^{-1}=Id_{Y}$ for
some $f^{-1}:Y\rightarrow X$ then $f$ is called a \textit{retraction}, and if
$f$ satisfies the ``left invertibility'' condition $f^{-1}\circ f=Id_{X}$ ,
then it is called a \textit{coretraction}, where $Id_{X}$ and $Id_{Y}$ are
identity mappings $Id_{X}:X\rightarrow X$ and $Id_{Y}:Y\rightarrow Y$ for
which $\forall x\in X,$ $Id_{X}\left(  x\right)  =x$ and $\forall y\in Y,$
$Id_{Y}\left(  y\right)  =y$. These requirements sometimes are very strong to
be fulfilled (see e.g. \cite{dav/rob,rob/cap,ara}). To obtain more weak
conditions one has to introduced the following ``regularity'' conditions

\begin{equation}
f\circ f_{in}^{\ast}\circ f=f,\label{3r}
\end{equation}

where $f_{in}^{\ast}$ is called an \textit{inner inverse} \cite{nashed}, and
such $f$ is called \textit{regular}. Similar ``reflexive regularity'' conditions

\begin{equation}
f_{out}^{\ast}\circ f\circ f_{out}^{\ast}=f_{out}^{\ast}\label{4r}
\end{equation}

defines an \textit{outer inverse} $f_{out}^{\ast}$. Notice that in general
$f_{in}^{\ast}\neq f_{out}^{\ast}\neq f^{-1}$ or it can be that $f^{-1}$ does
not exist at all. If $f_{in}^{\ast}$ is an inner inverse, then

\begin{equation}
f^{\ast}=f_{in}^{\ast}\circ f\circ f_{in}^{\ast}\label{fin}
\end{equation}
is always both inner and outer inverse or \textit{generalized inverse}
(quasi-inverse) \cite{rao/mit,dav/rob}, and so for any regular $f$ there
exists (need not be unique) $f^{\ast}$ from (\ref{fin}) for which both
regularity conditions (\ref{3r}) and (\ref{4r}) hold.

Let us consider a composition of two morphisms and its ``invertibility''
properties. It can be shown, that a retraction is an epimorphism, and a
regular monomorphism is a coretraction \cite{nashed}. If $f$ and $g$ are
regular, it is not necessary that their product $f\circ g$ is also regular,
but, if $f\circ g$ is regular and $g$ is an epimorphism, then $f$ is also
regular, and opposite, if $f\circ g$ is and $f$ is a monomorphism, then $g$ is regular.

If the composition $h=g\circ f$ belongs to the same class of functions
(closure), then all such morphisms form a semigroup of such functions
\cite{mag13,mag01,mag2}. If for any $f:X\rightarrow Y$ there will be a unique
$f^{\ast}:Y\rightarrow X$ satisfying (\ref{3r})--(\ref{4r}), this semigroup is
called an inverse semigroup \cite{petrich3} which we denote $\frak{F}$.

Let us define two idempotent ``projection operators'' $\mathcal{P} _{f}=f\circ
f^{\ast}$, $\mathcal{P}_{f}:Y\rightarrow Y$ and $\mathcal{P} _{f^{\ast}
}=f^{\ast}\circ f$, $\mathcal{P}_{f^{\ast}}:X\rightarrow X$ satisfying
$\mathcal{P}_{f}\circ\mathcal{P}_{f}=\mathcal{P}_{f}$, $\mathcal{P }_{f}\circ
f=f\circ\mathcal{P}_{f^{\ast}}=f$ and $\mathcal{P} _{f^{\ast}}\circ
\mathcal{P}_{f^{\ast}}=\mathcal{P}_{f^{\ast}}$, $\mathcal{P} _{f^{\ast}}\circ
f^{\ast}=f^{\ast}\circ\mathcal{P}_{f}=f^{\ast}$. If we introduce the $\ast
$-operation $\left(  f\right)  ^{\ast}=f^{\ast}$ by formulas (\ref{3r}
)--(\ref{4r}) and assume that this operation acts on the product of morphisms
$f:X\rightarrow Y$ and $g:Y\rightarrow Z$ in the following way $\left(  g\circ
f\right)  ^{\ast}=f^{\ast}\circ g^{\ast}$, then commutativity of projectors
$\mathcal{P}_{f}\circ\mathcal{P}_{g^{\ast}}= \mathcal{P}_{g^{\ast}}
\circ\mathcal{P}_{f}$ leads to closure of the semigroup, i.e. the product
$g\circ f$ also satisfies both regularity conditions (\ref{3r})--(\ref{4r}).

\section{Higher analogs of regularity}

Here we try to construct higher analogs of regularity conditions (\ref{3r}
)--(\ref{4r}). They were proposed for some particular case (noninvertible
analogs of supermanifolds) in \cite{dup14,dup18} (for other generalizations of
regularity see e.g. \cite{laj/sza,fri/mor}).

Let we have two elements $f$ and its regular $f^{\ast}$ (in sense of
(\ref{3r})) of semigroup $\frak{F}$. Consider a third morphism $f^{\ast\ast
}:X\rightarrow Y$ and analyze the action $f\circ f^{\ast}\circ f^{\ast\ast
}:X\rightarrow Y$. This means the composition $f\circ f^{\ast}\circ
f^{\ast\ast}$ cannot be equal to identity $Id_{X}$. Therefore it is possible
to ``regularize'' $f\circ f^{\ast}\circ f^{\ast\ast}$ in the following way

\begin{equation}
f\circ f^{\ast}\circ f^{\ast\ast}\circ f^{\ast}=f^{\ast}.\label{ffff}
\end{equation}

This formula can be called as 2-regularity condition and be considered as a
definition of $\ast\ast$-operation. For 3-regularity and $f^{\ast\ast\ast
}:Y\rightarrow X$ we can obtain an analog of (\ref{3r}) in the form

\begin{equation}
f\circ f^{\ast}\circ f^{\ast\ast}\circ f^{\ast\ast\ast}\circ f=f.\label{3f}
\end{equation}

By recursive considerations we can propose the following formula of $n$ -regularity

\begin{align}
f\circ f^{\ast}\circ f^{\ast\ast}\ldots\circ f^{\overset{2k}{\overbrace
{\ast\ast\ldots\ast}}}\circ f^{\ast}  &  =f^{\ast}\label{nreg1}\\
f\circ f^{\ast}\circ f^{\ast\ast}\ldots\circ f^{\overset{2k+1}{\overbrace
{\ast\ast\ldots\ast}}}\circ f  &  =f.\label{nreg2}
\end{align}

Note that for even number of stars $f^{\overset{2k}{\overbrace{\ast\ast
\ldots\ast}}}:X\rightarrow Y$ and for odd number of stars $f^{\overset
{2k+1}{\overbrace{\ast\ast\ldots\ast}}}:Y\rightarrow X$. In general case all
$f^{\overset{n}{\overbrace{\ast\ast\ldots\ast}}}$ are different, and, for
instance, $\left(  f^{\ast}\right)  ^{\ast}\neq f^{\ast\ast}$. The action of
$\overset{n}{\overbrace{\ast\ast\ldots\ast}}$-operation on product depends
from number of stars and is the following

\begin{align}
\left(  g\circ f\right)  ^{\overset{2k+1}{\overbrace{\ast\ast\ldots\ast}}}  &
=f^{\overset{2k+1}{\overbrace{\ast\ast\ldots\ast}}}\circ g^{\overset
{2k+1}{\overbrace{\ast\ast\ldots\ast}}},\label{gf1}\\
\left(  g\circ f\right)  ^{\overset{2k}{\overbrace{\ast\ast\ldots\ast}}}  &
=g^{\overset{2k}{\overbrace{\ast\ast\ldots\ast}}}\circ f^{\overset
{2k}{\overbrace{\ast\ast\ldots\ast}}}.\label{gf2}
\end{align}

We can introduce ``higher projector'' by the formula
\begin{equation}
\mathcal{P}_{f}^{\left(  n\right)  }=f\circ f^{\ast}\circ f^{\ast\ast}
\ldots\circ f^{\overset{n}{\overbrace{\ast\ast\ldots\ast}}}.\label{pf}
\end{equation}

It is easy to check the following properties
\begin{align}
\mathcal{P}_{f}^{\left(  2k\right)  }\circ f^{\ast}  &  =f^{\ast}
,\label{pff1}\\
\mathcal{P}_{f}^{\left(  2k+1\right)  }\circ f  &  =f.\label{pff2}
\end{align}

and idempotence $\mathcal{P}_{f}^{\left(  n\right)  }\circ\mathcal{P}
_{f}^{\left(  n\right)  }=\mathcal{P}_{f}^{\left(  n\right)  }$.

\section{Semicommutative diagrams and obstruction}

In previous subsection we considered morphisms and regularity properties for
two given objects $X$ and $Y$, because we studied various types of inverses.
Now we will extend these consideration to any number of objects and introduce
semicommutative diagrams (firstly considered in \cite{dup14,dup18} ).

Obviously, that for two morphisms $f:X\rightarrow Y$ and $g:Y\rightarrow X$
instead of ``invertibility'' $g\circ f=Id_{X}$ we have the same generalization
as regularity (\ref{3r}), i.e. $f\circ g\circ f=f$, where $g$ plays the role
of an inner inverse \cite{nashed}. \medskip

\begin{center}
\setlength{\unitlength}{.3in} \begin{picture}(15,3.5)
\put(4.6,3.2){{ $f$}}
\put(4.6,1.7){{ $g$}}
\put(4,2.8){\vector(1,0){2.1}}
\put(6.1,2.4){\vector(-1,0){2.1}}
\put(6.3,3.4){\small``Regularization''}
\put(8,2.5){$\Longrightarrow$}
\put(10.6,1.6){{ $g$}}
\put(10.6,3.2){{ $f$}}
\put(10,2.9){\vector(1,0){2.1}}
\put(12.1,2.3){\vector(-1,0){2.1}}
\put(10,2.6){\vector(1,0){2.1}}
\put(0,2.5){\large$n=2$}
\put(4,0.2){\makebox(1,1){\small Invertible morphisms}}
\put(12,0.2){\makebox(1,1){\small Noninvertible (regular)
morphisms}}
\end{picture}
\end{center}

Usually, for 3 objects $X,Y,Z$ and 3 morphisms $f:X\rightarrow Y$ and
$g:Y\rightarrow Z$ and $h:Z\rightarrow X$ one can have the ``invertible''
triangle commutative diagram $h\circ g\circ f=Id_{X}$. Its regular extension
has the form

\begin{equation}
f\circ h\circ g\circ f=f.\label{fgh}
\end{equation}
Such a diagram (from the right)

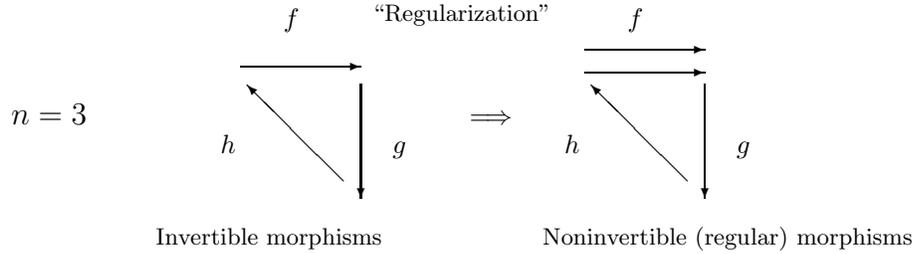
\begin{center}
\setlength{\unitlength}{.3in} \begin{picture}(15,5)
\put(5.8,1.5){\vector(-1,1){1.68}}
\put(4.6,4.2){{ $f$}}
\put(3.5,2){{ $h$}}
\put(4,3.5){\vector(1,0){2.1}}
\put(6.1,3.2){\vector(0,-1){2}}
\put(6.3,4.3){\small``Regularization''}
\put(8,2.5){$\Longrightarrow$}
\put(6.5,2){{ $g$}}
\put(9.5,2){{ $h$}}
\put(11.8,1.5){\vector(-1,1){1.68}}
\put(10.6,4.2){{ $f$}}
\put(10,3.8){\vector(1,0){2.1}}
\put(10,3.4){\vector(1,0){2.1}}
\put(12.1,3.2){\vector(0,-1){2}}
\put(12.5,2){{ $g$}}
\put(0,2.5){\large$n=3$}
\put(4,0){\makebox(1,1){\small Invertible morphisms}}
\put(12,0){\makebox(1,1){\small Noninvertible (regular)
morphisms}}
\end{picture}
\end{center}

\noindent can be called a \textit{semicommutative diagram}. This triangle case
can be expanded on any number of objects and morphisms.

To measure difference between semicommutative and commutative cases let us
introduce self-mappings $e_{X}^{\left(  n\right)  }:X\rightarrow X$ which are
defined by

\begin{align}
e_{X}^{\left(  1\right)  }  &  =Id_{X},\label{e1}\\
e_{X}^{\left(  2\right)  }  &  =g\circ f,\label{e2}\\
e_{X}^{\left(  3\right)  }  &  =h\circ g\circ f,\label{e3}\\
&  \ldots\nonumber
\end{align}

It is obvious that for commutative diagrams all $e_{X}^{\left(  n\right)  }$
are equal to identity $e_{X}^{\left(  n\right)  }=Id_{X}$. The deviation of
$e_{X}^{\left(  n\right)  }$ from identity will give us measure of obstruction
of commutativity, and therefore we call $e_{X}^{\left(  n\right)  }$ \textit{
obstructors}. The minimum number $n=n_{obstr}$ for which $e_{X}^{\left(
n\right)  }\neq Id_{X}$ occurs will define a quantitative measure of
obstruction $n_{obstr}$.

In terms of obstructors $e_{X}^{\left(  n\right)  }$ the $n$-regularity
condition can be written in the short form

\begin{equation}
f\circ e_{X}^{\left(  n\right)  }=f.\label{ef}
\end{equation}

From definitions (\ref{e1})--(\ref{e3}) and (\ref{ef}) it simply follows that
obstructors $e_{X}^{\left(  n\right)  }$ are idempotents.

It can be noted that for a given $n$ additional arrows do not lead to new
commutativity relations due to (\ref{ef}). Therefore only one additional arrow
can be important for extension of the commutativity to semicommutativity.

\section{Monoidal categories, functors and regularity}

The morphisms $e_{X}^{\left(  n\right)  }$ can be used to extend the notion of
a functor $\mathsf{F}:\frak{C}_{1}\rightarrow\frak{C}_{2}$. All the standard
definitions of a functor (as a mapping of one category to another with
preserving composition of morphisms \cite{mitchell,maclane1}) do not changed,
but preservation of identity $\mathsf{F}\left(  Id_{X_{1}}\right)  =Id_{X_{2}
}$, where $X_{2}=\mathsf{F}X_{1}$, $X_{1}\in\mathrm{Ob}\frak{C} _{1}$,
$X_{2}\in\mathrm{Ob}\frak{C}_{2}$, can be replaced by requirement of
preservation of morphisms $e_{X}^{\left(  n\right)  }$ as

\begin{equation}
\mathsf{F}^{\left(  n\right)  }\left(  e_{X_{1}}^{\left(  n\right)  }\right)
=e_{X_{2}}^{\left(  n\right)  },\label{fe}
\end{equation}
where $e_{X_{1}}^{\left(  n\right)  }\in\mathrm{Mor}\frak{C}_{1}$, $e_{X_{2}
}^{\left(  n\right)  }\in\mathrm{Mor}\frak{C}_{2}$ defined in (\ref{e1}
)--(\ref{e3}) for two categories. Then the generalized functor $\mathsf{F}
^{\left(  n\right)  }$ becomes $n$-dependent. From (\ref{e1}) it follows that
$n=1$ corresponds to the standard functor, i.e. $\mathsf{F}^{\left(  1\right)
}=\mathsf{F}$. In the same manner we can ``regularize'' natural
transformations \cite{mitchell}.

Let $\frak{C}$ be a monoidal category \cite{mitchell,maclane1} equipped with a
monoidal operation $\otimes:\frak{C}\times\frak{C}\longrightarrow\frak{C }$. A
triple of objects $X,Y,Z$ is said to be a regular $3$-cycle if and only if
every sequence of morphisms $X\overset{f}{\longrightarrow}Y\overset{
g}{\longrightarrow}Z\overset{h}{\longrightarrow}X$ define uniquely the
morphism $e_{X}^{\left(  3\right)  }:X\longrightarrow X$ by the following
relation $e_{X}^{\left(  3\right)  }:=h\circ g\circ f$ and subjects to the
relation $f\circ h\circ g\circ f=f$. The object $Y$ is said to be a \textit{
(first) regular dual} of $X$, and the object $Z$ is called the \textit{ second
regular dual} of $X$. We denote by $C_{3}\left(  \frak{C}\right)  $ the
collection of all regular $3$-cycles on $\frak{C}$ . This collection is said
to be \textit{regularity} in $\frak{C}$. The generalization to arbitrary
$n\geq4$ is obvious. Let $X,Y,Z$ and $X^{\prime},Y^{\prime},Z^{\prime}$ be two
regular $3$-cycles in $\frak{C}$. Then the morphism $f:X\longrightarrow
X^{\prime}$ such that $f\circ e_{X}^{\left(  3\right)  }=e_{X^{\prime}
}^{\left(  3\right)  }\circ f$ is said to be a \textit{3-cycle morphism}. If
$f:X\longrightarrow X^{\prime}$ and $g:X^{\prime}\longrightarrow
X^{\prime\prime}$ are two 3-cycle morphisms, then the composition $g\circ
f:X\rightarrow X^{\prime\prime}$ is also a 3-cycle morphism. Moreover the
regularity $C_{3}\left(  \frak{C}\right)  $ forms a monoidal category with
3-cycles as objects and 3-cycle morphisms. The monoidal product of two regular
3-triples $X,Y,Z$ and $X^{\prime},Y^{\prime},Z^{\prime}$ is the triple
$X\otimes X^{\prime},Y\otimes Y^{\prime},Z\otimes Z^{\prime}$ which is also a
regular $3$-cycle. The category $C_{3}\left(  \frak{C}\right)  $ is said to be
a \textit{regularization} of $\frak{C}$ .

Let us consider a symmetric monoidal category $\frak{C}$
\cite{maclane1,joy/str} playing an important role in topological QFT
\cite{bae/dol}, supersymmetry \cite{lyu}, quantum groups
\cite{maj,yet,maj1,lyu1} and quantum statistics \cite{marc0,marc1,marc2}.In
$\frak{C}$ for any two objects $X$ and $Y$ and the operation $X\otimes Y$ one
usually defines a natural isomorphism (``braiding'' \cite{joy/str}) by
$\mathrm{B} _{X,Y}:X\otimes Y\rightarrow Y\otimes X$ satisfying the symmetry
condition (``invertibility'')

\begin{equation}
\mathrm{B}_{Y,X}\circ\mathrm{B}_{X,Y}=Id_{X\otimes Y}\label{bbi}
\end{equation}
which formally defines $\mathrm{B}_{Y,X}=\mathrm{B}_{X,Y}^{-1}:Y\otimes
X\rightarrow X\otimes Y$. Note that possible nonsymmetric braiding in context
of the noncommutative geometry was considered in \cite{marc3,maj3} (see also
\cite{gre/ser}). Here we introduce a ``regular'' extension of the symmetry
condition (\ref{bbi}) in the form

\begin{equation}
\mathrm{B}_{X,Y}\circ\mathrm{B}_{X,Y}^{\ast}\circ\mathrm{B}_{X,Y} =\mathrm{B}
_{X,Y},\label{bbb}
\end{equation}
where in general $\mathrm{B}_{X,Y}^{\ast}\neq\mathrm{B}_{X,Y}^{-1}$. Such a
category can be called a ``regular'' category to distinct from symmetric and
``braided'' categories \cite{maclane1,joy/str}.

In categorical sense the prebraiding relations usually are defined as
\cite{marc1,maj2,joy/str}
\begin{align}
\mathrm{B}_{X\otimes Y,Z}  &  =\mathbf{B}_{X,Z,Y}^{R}\circ\mathbf{B}
_{X,Y,Z}^{L},\label{bb1}\\
\mathrm{B}_{Z,X\otimes Y}  &  =\mathbf{B}_{X,Z,Y}^{L}\circ\mathbf{B}
_{X,Y,Z}^{R},\label{bb2}\\
\mathbf{B}_{X,Y,Z}^{L}  &  =Id_{X}\otimes\mathrm{B}_{Y,Z},\label{bb3}\\
\mathbf{B}_{X,Y,Z}^{R}  &  =\mathrm{B}_{X,Y}\otimes Id_{Z},\label{bb4}
\end{align}
and prebraidings $\mathrm{B}_{X\otimes Y,Z}$ and $\mathrm{B}_{Z,X\otimes Y}$
satisfy (for symmetric case) the ``invertibility'' property $\mathrm{B}
_{X\otimes Y,Z}^{-1}\circ\mathrm{B}_{X\otimes Y,Z}=Id_{X\otimes Y\otimes Z}$ ,
where $\mathrm{B}_{X\otimes Y,Z}^{-1}=\mathrm{B}_{Z,X\otimes Y}$. In this
notations the standard ``invertible'' Yang-Baxter equation is \cite{maj2,maj3}

\begin{equation}
\mathbf{B}_{Y,Z,X}^{R}\circ\mathbf{B}_{Y,X,Z}^{L}\circ\mathbf{B}_{X,Y,Z}^{R}=
\mathbf{B}_{Z,X,Y}^{L}\circ\mathbf{B}_{X,Z,Y}^{R}\circ\mathbf{B}_{X,Y,Z}
^{L}.\label{yb0}
\end{equation}

Possible ``noninvertible'' (endomorphism semigroup) solutions of this equation
without introduction of $e_{X}^{\left(  n\right)  }$ were studied in
\cite{fangli3,fangli2,fangli1}. For ``noninvertible'' braidings satisfying
regularity (\ref{bbb}) it is naturally to exploit the obstructors
$e_{X}^{\left(  n\right)  }$ instead of identity $Id_{X}$ as

\begin{align}
\mathbf{B}_{X,Y,Z}^{L\left(  n\right)  }  &  =e_{X}^{\left(  n\right)
}\otimes\mathrm{B}_{Y,Z},\label{eb1}\\
\mathbf{B}_{X,Y,Z}^{R\left(  n\right)  }  &  =\mathrm{B}_{X,Y}\otimes
e_{Z}^{\left(  n\right)  },\label{eb2}
\end{align}
to weaken prebraiding construction in the following way
\begin{align}
\mathrm{B}_{X\otimes Y,Z}^{\left(  n\right)  }  &  =\mathbf{B}_{X,Z,Y}
^{R\left(  n\right)  }\circ\mathbf{B}_{X,Y,Z}^{L\left(  n\right)
},\label{rb1}\\
\mathrm{B}_{Z,X\otimes Y}^{\left(  n\right)  }  &  =\mathbf{B}_{X,Z,Y}
^{L\left(  n\right)  }\circ\mathbf{B}_{X,Y,Z}^{R\left(  n\right)
},\label{rb2}
\end{align}

Then their ``invertibility'' can be also ``regularized'' as follows
\begin{equation}
\mathrm{B}_{X\otimes Y,Z}^{\left(  n\right)  }\circ\mathrm{B}_{X\otimes
Y,Z}^{\left(  n\right)  \ast}\circ\mathrm{B}_{X\otimes Y,Z}^{\left(  n\right)
}= \mathrm{B}_{X\otimes Y,Z}^{\left(  n\right)  },\label{bbbb}
\end{equation}

where in general case $\mathrm{B}_{X\otimes Y,Z}^{\left(  n\right)  \ast}
\neq\mathrm{B}_{X\otimes Y,Z}^{-1}$. Thus the corresponding $n$
-``noninvertible'' analog of the Yang-Baxter equation (\ref{rb2}) is
\begin{equation}
\mathbf{B}_{Y,Z,X}^{R\left(  n\right)  }\circ\mathbf{B}_{Y,X,Z}^{L\left(
n\right)  }\circ\mathbf{B}_{X,Y,Z}^{R\left(  n\right)  }=\mathbf{B}
_{Z,X,Y}^{L\left(  n\right)  }\circ\mathbf{B}_{X,Z,Y}^{R\left(  n\right)
}\circ\mathbf{B}_{X,Y,Z}^{L\left(  n\right)  }.\label{yb1}
\end{equation}

Its solutions can be found by application of the semigroup methods (see e.g.
\cite{fangli2,fangli3}).

The introduced formalism can be used in analysis of categories with some
weaken invertibility conditions, which can appear in nontrivial supersymmetric
or noncommutative geometry constructions beyond the group theory.

\bigskip

\textbf{Acknowledgments}. One of the authors (S.D.) would like to thank Jerzy
Lukierski for kind hospitality at the University of Wroc\l aw, where this work
was done, and to Andrzej Borowiec, Andrzej Frydryszak and Cezary Juszczak for
valuable discussions and help during his stay in Wroc\l aw.


\newpage
\begin{thebibliography}{99}
\bibitem{neu}J.~von Neumann, \newblock"On regular rings," \newblock Proc. Nat.
Acad. Sci. USA \newblock{\bf22}, 707 (1936).

\bibitem{pen1}R.~Penrose, \newblock"A generalized inverse for matrices,"
\newblock Math. Proc. Cambridge Phil. Soc. \newblock{\bf51}, 406 (1955).

\bibitem{rabson}G.~Rabson, \newblock{\it The Generalized Inverses in Set
Theory and Matrix Theory} \newblock(Amer. Math. Soc., Providence, 1969).

\bibitem{rao/mit}C.~R. Rao and S.~K. Mitra, \newblock{\it Generalized
Inverse of Matrices and its Application} \newblock(Wiley, New York, 1971).

\bibitem{nashed}M.~Z. Nashed, \newblock{\it Generalized Inverses and
Applications} \newblock(Academic Press, New York, 1976).

\bibitem{mun/pen}W.~D. Munn and R.~Penrose, \newblock"Pseudoinverses in
semigroups," \newblock Math. Proc. Cambridge Phil. Soc. \newblock{\bf57}, 247 (1961).

\bibitem{cli5}A.~H. Cliford, \newblock"The fundamental representation of a
regular semigroup," \newblock Semigroup Forum \newblock{\bf10}, 84 (1975/76).

\bibitem{lal1}G.~Lallement, \newblock"Structure theorems for regular
semigroups," \newblock Semigroup Forum \newblock{\bf4}, 95 (1972).

\bibitem{howie}J.~M. Howie, \newblock{\it An Introduction to Semigroup
Theory} \newblock(Academic Press, London, 1976).

\bibitem{lawson}M.~V. Lawson, \newblock{\it Inverse Semigroups: {T}he
Theory of Partial Symmetries} \newblock(World Sci., Singapore, 1998).

\bibitem{mitchell}B.~Mitchell, \newblock{\it Theory of Categories}
\newblock(Academic Press, New York, 1965).

\bibitem{maclane1}S.~MacLane, \newblock{\it Categories for the Working
Mathematician} \newblock(Springer-Verlag, Berlin, 1971).

\bibitem{dup14}S.~Duplij, \newblock"Noninvertibility and ''semi-'' analogs of
(super) manifolds, fiber bundles and homotopies,", Univ. Kaiserslautern
\textit{preprint}, KL-TH-96/10, \texttt{q-alg/9609022}, Kaiserslautern, 1996.

\bibitem{dup18}S.~Duplij, \newblock"On semi-supermanifolds," \newblock
Pure Math. Appl. \newblock{\bf9}, 1 (1998).

\bibitem{cohn}P.~M. Cohn, \newblock{\it Universal Algebra} \newblock
(Harper \& Row, New York, 1965).

\bibitem{dup6}S.~Duplij, \newblock"On semigroup nature of superconformal
symmetry," \newblock J.~Math. Phys. \newblock{\bf32}, 2959 (1991).

\bibitem{dup11}S.~Duplij, \newblock"Some abstract properties of semigroups
appearing in superconformal theories," \newblock Semigroup Forum \newblock
{\bf54}, 253 (1997).

\bibitem{dup-hab}S.~Duplij, \newblock{\it
Semigroup methods in supersymmetric theories of elementary
particles} \newblock(Habilitation Thesis, Kharkov State University,
\texttt{\ math-ph/9910045}, Kharkov, 1999).

\bibitem{petrich3}M.~Petrich, \newblock{\it Inverse Semigroups} \newblock
(Wiley, New York, 1984).

\bibitem{dav/rob}D.~L. Davis and D.~W. Robinson, \newblock"Generalized
inverses of morphisms," \newblock Linear Algebra Appl. \newblock{\bf5}, 329 (1972).

\bibitem{rob/cap}J.~A.~G. Roberts and H.~W. Capel, \newblock"Area preserving
mappings that are not reversible," \newblock Phys. Lett. \newblock{\bf A162},
243 (1992).

\bibitem{ara}A.~Arai, \newblock"Noninvertible {B}ogolyubov transformations and
instability of embedded eigenvalues," \newblock J. Math. Phys. \newblock
{\bf32}, 1838 (1991).

\bibitem{mag13}K.~D. Magill, \newblock"Semigroups of functions on topological
spaces," \newblock Proc. London Math. Soc. \newblock{\bf16}, 507 (1966).

\bibitem{mag01}K.~D. Magill, \newblock"Isotopisms of semigroups of functions,"
\newblock Trans. Amer. Math. Soc. \newblock{\bf148}, 121 (1970).

\bibitem{mag2}K.~D. Magill, \newblock"Some open problems and directions for
further research in semigroups of continuous selfmaps," \newblock in
\textit{Universal algebra and applications}, \newblock Number~9 in Banach
Center Publications \newblock(PWN --- Polish Scientific Publishers, Warsaw,
1982), \newblock p. 439.

\bibitem{laj/sza}S.~Lajos and G.~Sz\'{a}sz, \newblock"Generalized regularity
in semigroups,", Karl Marx Univ. of Economics \textit{preprint}, 1975-7,
Budapest, 1975.

\bibitem{fri/mor}R.~Friedman and J.~W. Morgan, \newblock"Obstruction bundles,
semiregularity, and {S}eiberg-{W}itten,", Columbia Univ. \textit{ preprint},
\texttt{alg-geom/9509007}, New York, 1995.

\bibitem{joy/str}A.~Joyal and R.~Street, \newblock"Braided monoidal
categories,", Macquarie University \textit{preprint}, Mathematics Reports
86008, North Ryde, New South Wales, 1986.

\bibitem{bae/dol}J.~C. Baez and J.~Dolan, \newblock"Higher-dimensional algebra
and topological quantum field theory," \newblock J.~Math. Phys. \newblock
{\bf36}, 6073 (1995).

\bibitem{lyu}V.~V. Lyubashenko, \newblock"The {B}erezinian in some monoidal
categories," \newblock Ukrainian Math. J. \newblock{\bf38}, 588 (1986).

\bibitem{maj}S.~Majid, \newblock"Braided groups and duals of monoidal
categories,", Univ. Cambridge \textit{preprint}, DAMTP-92-03, Cambridge, 1992.

\bibitem{yet}D.~N. Yetter, \newblock"Quantum groups and representations of
monoidal categories," \newblock Math. Proc. Camb. Phil. Soc. \newblock
{\bf108}, 261 (1990).

\bibitem{maj1}S.~Majid, \newblock"Duals and doubles of monoidal categories,",
Univ. Cambridge \textit{\ preprint}, DAMTP-89/41, Cambridge, 1989.

\bibitem{lyu1}V.~V. Lyubashenko, \newblock"Tangles and {H}opf algebras in
braided categories," \newblock J. Pure and Appl. Algebra \newblock{\bf98} ,
245 (1995).

\bibitem{marc0}W.~Marcinek, \newblock"On braid statistics and noncommutative
calculus," \newblock Rep. Math. Phys. \newblock{\bf33}, 117 (1993).

\bibitem{marc1}W.~Marcinek, \newblock"Categories and quantum statistics,"
\newblock Rep. Math. Phys. \newblock{\bf38}, 149 (1996).

\bibitem{marc2}W.~Marcinek, \newblock"On unital braiding and quantization,"
\newblock Rep. Math. Phys. \newblock{\bf34}, 325 (1994).

\bibitem{marc3}W.~Marcinek, \newblock"Noncommutative geometry for arbitrary
braidings," \newblock J. Math. Phys. \newblock{\bf35}, 2633 (1996).

\bibitem{maj3}S.~Majid, \newblock"Quasitriangular {H}opf algebras and {Y}
ang-{B}axter equations," \newblock Int. J. Mod. Phys. \newblock{\bf A5}, 1 (1990).

\bibitem{gre/ser}P.~Greenberg and V.~Sergiescu, \newblock"An acyclic extension
of the braid group," \newblock Comm. Math. Helv. \newblock{\bf
62}, 185 (1991).

\bibitem{maj2}S.~Majid, \newblock"Solutions of the {Y}ang-{B}axter equations
from braided-{L}ie algebras and braided groups," \newblock J. Knot Theor.
Ramifications \newblock{\bf4}, 673 (1995).

\bibitem{fangli3}F.~Li, \newblock"Weak {H}opf algebras and new solutions of
{Y}ang-{B}axter equation," \newblock J. Algebra \newblock{\bf208}, 72 (1998).

\bibitem{fangli2}F.~Li, \newblock"Solutions of {Y}ang-{B}axter equation in an
endomorphism semigroup and quasi-(co)braided almost bialgebras,", Zhejiang
Univ. \textit{preprint}, Hangzhou, 1999.

\bibitem{fangli1}F.~Li, \newblock"Weaker structures of {H}opf algebras and
singular solutions of {Y}ang-{B}axter equation,", Zhejiang Univ. \textit{
preprint}, Hangzhou, 2000.
\end{thebibliography}
\end{document}